\documentclass{article}
\usepackage{graphicx} 
\usepackage{float} 
\usepackage[sorting=none,maxbibnames=3,giveninits=true,doi=false,isbn=false,url=false]{biblatex} 
\usepackage{authblk}  
\usepackage{anyfontsize}
\usepackage{xcolor}
\usepackage{makecell}

\title{Feasibility study of a mission to Sedna - Nuclear propulsion and advanced solar sailing concepts}

\author[1]{Elena Ancona}
\author[2]{Roman Ya. Kezerashvili}
\author[3]{Savino Longo}
\affil[1]{Department of Mechanics, Mathematics \& Management, Politecnico di Bari, 70126 Bari, Italy, \textit{elena.ancona@poliba.it}}
\affil[2]{Physics Department, New York City College of Technology, CUNY, Brooklyn, 11201 NY, USA, \textit{rkezerashvili@citytech.cuny.edu}}
\affil[3]{Dipartimento di Chimica, Università degli Studi di Bari, 70126 Bari, Italy}

\date{\today}

\begin{document}

\maketitle

\begin{abstract}

Following the discovery of Pluto, interest in exploring trans-Neptunian objects has grown significantly. This work examines mission opportunities to reach the outermost regions of the Solar System, focusing on the dwarf planet Sedna, discovered in 2003.

Exploring the outer reaches of the Solar System presents significant propulsion and mission design challenges. This study assesses the feasibility of a mission to Sedna using two advanced propulsion concepts: the Direct Fusion Drive (DFD) rocket engine, based on D-$^{3}$He thermonuclear fusion, and a solar sail utilizing thermal desorption of its coating for propulsion. Both are evaluated for a one-way Earth-to-Sedna mission; however, due to the different performances the DFD would enable orbit insertion, whereas for the solar sail a flyby is envisioned. The analysis evaluates key mission parameters, including delivered payload capacity, travel time, and potential science return. For the DFD, we assume a 1.6 MW system with constant thrust and specific impulse, while for the solar sail, we consider acceleration via thermal desorption and a gravity-assist maneuver around Jupiter. The mission analysis incorporates four key phases: departure, interplanetary acceleration, interplanetary coasting, and rendezvous. Sedna is expected to pass through the perihelion of its orbit in 2075–2076 and then move again away from the Sun. Considering the distances involved, a mission targeting the object would need to be launched "relatively" soon, especially if using conventional propulsion systems, that could require up to 30 years of deep-space travel. In our study, results indicate that the DFD could reach Sedna in approximately 10 years, with 1.5 years of thrusting, while the solar sail, assisted by Jupiter’s gravity, could complete the journey in 7 years. The feasibility of science payload accommodation, power availability, and communication constraints is also considered. These findings provide a comparative foundation for future deep-space mission planning.

\vspace{0.2cm}


\end{abstract}

\maketitle

\section{Introduction}

Exploration missions targeting the Kuiper Belt, the Oort Cloud, the gravitational focal point of the Sun, and even extending to the Alpha-Centauri system represent the next frontier in space exploration. Among potential destinations, there is a growing focus within the scientific community on the trans-Neptunian object Sedna (90377) discovered in 2003 \cite{BrownSedna}. With its highly eccentric orbit around the Sun, Sedna is currently approaching its perihelion at around 76 AU, offering an exceptional opportunity for deep-space exploration. Its aphelion, situated approximately 936 AU away, makes Sedna an intriguing target for further study. Given its orbital period of 11 thousand years, scientists have been proposing missions for launch in the next few years (around 2030), including gravity assist that would allow reaching Sedna in time for its closest approach in 2075 \cite{Zubko2021}. 
Considered methods include chemical propulsion with gravity assist maneuvers near Venus, Earth, Jupiter, and Neptune, as well as an Oberth maneuver near the Sun \cite{Zubko2022}. In this work we present possibilities to explore Sedna considering missions with alternative propulsion: using a thermonuclear Direct Fusion Drive (DFD) and solar sail with thermal desorption of the coating. 

The Direct Fusion Drive rocket engine is under development at Princeton University Plasma Physics Laboratory \cite{Cohen2000,Cohen2019}. The underlying physics for the Princeton Field-Reversed Configuration (PFRC) and Direct Fusion Drive have been previously described in  \cite{Cohen2000,Cohen2019,Cohen2007,Paluszek2023}. 
A review article \cite{Cohen2023} on the PFRC reactor concept includes the status of development, the proposed path toward a reactor, and the commercialization potential of a PFRC reactor. 
The DFD is the D-$^{3}$He-fueled, aneutronic, thermonuclear fusion propulsion thruster that employs a unique plasma heating system to produce nuclear fusion power in the range of 1 to 10 MW. Missions design using DFD were considered in \cite{Genta2020,Gajeri2021,Aime2021,KezerashviliDFD2021,Paluszek2023}. The DFD presents a promising alternative to conventional propulsion, offering high thrust-to-weight ratio and continuous acceleration. However, its feasibility remains subject to key engineering challenges, including plasma stability, heat dissipation, and operational longevity under deep-space radiation. While advances in fusion-based propulsion are being made, a Sedna-class mission requires evaluating whether current DFD designs can sustain long-duration operations and provide reliable power for onboard instruments; this could be explored in future research.

Solar sails, first theorized by K. Tsiolkovsky and F. Tsander back in the 1920s, are large sheets of low-areal density material whose only source of energy is the Sun photons' flux \cite{tsander1969scientific}. At least in theory, a solar sail mission could be of unlimited duration, thanks to the “ever-present gentle push of sunlight” \cite{Vulpetti2008}; a remarkable advantage is that no propellant is needed. The idea of taking advantage of thermal desorption of the coating for a solar sail was generalized in \cite{Kezer2015} considering a solar sail coated by materials that undergo desorption at a particular temperature as a result of heating by the solar electromagnetic radiation at a particular heliocentric distance. 
While solar sailing has been extensively studied for deep-space applications, its feasibility for a Sedna mission requires assessment in terms of long-duration structural integrity, propulsion efficiency, and power availability for science operations. The recent NASA Advanced Composite Solar Sail System (ACS3) demonstration validated critical aspects of sail deployment and control \cite{ACS3_Smallsat, ACS3news}, but extending this technology to interstellar precursor missions introduces challenges related to durability, navigation precision, and long-term trajectory optimization.

This paper presents a preliminary feasibility assessment of a mission to Sedna, focusing on two advanced propulsion concepts: Direct Fusion Drive (DFD) and augmented solar sails. The study aims to evaluate these architectures based on key mission requirements, including travel time, payload capacity, power availability, and communication constraints. Rather than providing a fully optimized mission design, this work explores the trade-offs and constraints associated with each approach, identifying the critical challenges and feasibility boundaries. The analysis includes trajectory considerations, propulsion system constraints, and an initial assessment of science payload accommodation. By structuring the feasibility assessment across these categories, this study provides a foundation for future, more detailed mission designs.

This study defines key functional requirements for a Sedna mission, including propulsion constraints, power generation for scientific instruments, and communication feasibility. The following sections outline these considerations in detail, evaluating the trade-offs between DFD and augmented solar sail architectures.

The article is organized in the following way: in Sec. 2 we describe the motivation to explore the trans-Neptunian object Sedna, the main challenges, and missions considered in literature based on chemical propulsion. Concepts of propulsion using the thermonuclear fusion based on the DFD fusion rocket engine and solar sail with thermal desorption of the coating are presented in Sec. 3. Proposed scenarios of the mission profiles for both the DFD rocket and the solar sail are given in Sec. 4, followed by the conclusions in Sect. 5.

\section{Mission to Sedna: a deep dive into the solar system}
Since the early days of space exploration, bodies like the Moon, Mars, Venus, and Jupiter have been popular targets due to their proximity, dimension, and similarities to Earth. However, the interest in more distant objects, like comets, asteroids, and icy moons, has been growing as technology improves and the quest to understand planetary formation and the potential for life beyond Earth continues. Together with Saturn's moon Enceladus \cite{Tsou2012} and Jupiter's moon Europa \cite{Lingam2023}, Sedna, in the inner Oort Cloud, is also being considered one of the most promising destinations \cite{BrownSedna2}. If considering a flyby mission, remote sensing instruments such as a high-resolution spectrometer and a wide-angle camera could provide insights into surface composition and geology, while a magnetometer could investigate Sedna’s interaction with the solar wind. However, precise gravitational field characterization, as done for planetary orbiters, would not be possible. This limitation is similar to that of New Horizons at Pluto, where only partial surface mapping and atmospheric escape measurements were achievable. In fact, due to the flyby nature of the encounter, New Horizons lacked the prolonged proximity and multiple orbital passes required to characterize higher-order gravitational harmonics. As a result, only the system's mass and limited geophysical data were retrieved \cite{stern2015pluto, mckinnon2017origin}. These constraints are intrinsic to flyby missions and stem from short interaction time, limited tracking data, and distance from the body, similar to the mission proposed.

This study evaluates Sedna mission feasibility at a systems level, analyzing propulsion performance, power availability, and communication constraints. While spacecraft subsystems are considered, this is not an in-depth engineering design of instruments or spacecraft hardware. The study outputs include trajectory feasibility, payload capability, and mission trade-offs.

\subsection{Reasons to explore the trans-Neptunian object Sedna}
A space mission to Sedna, a trans-Neptunian object (TNO) with a highly eccentric orbit, would be of great interest to the scientific community. Sedna's orbit takes it far beyond the Kuiper Belt, with its perihelion at around 76 AU and its aphelion at nearly 936 AU (31 times farther than Neptune's aphelion). Sedna could provide the first direct observations of this distant and unexplored region, which is crucial for understanding the outer solar system and its boundaries.
Its highly elongated orbit takes it well beyond the heliopause (situated at about 123 AU), the boundary beyond which the influence of the Sun is negligible with respect to that of the particles from interstellar space \cite{VoyagerPioneer}. This would have allowed Sedna to interact differently with other elements that could offer clues about the early history of the solar system, particularly the processes that shaped the orbits of such distant objects. 
Moreover, the fact that Sedna spends so little time in the proximity of the Sun (approximately 200–300 years within 100 AU, thus less than 3\% of its orbital period), protects its surface from solar radiation and heat that affect the surface composition \cite{SednaStrac}. In fact, spectroscopic observations suggest that it could contain organic compounds, water ice, and possibly methane ice \cite{OrganicSedna}. Analyzing its composition could reveal much about the original "building blocks" of planets and other bodies. In particular, a detailed analysis of any organic matter present on the surface of Sedna could provide specific information about the role of photochemical processes in the synthesis of organic matter versus alternative processes induced by fast electrons or other causes. This information, complementary to that which can be obtained from the analysis of Tholins for example on Europa or Titan, where photochemical processes are certainly more relevant, is of extreme interest in the current debate regarding the primary chemical reactions at the origin of life \cite{Shaw2007, Micca2021}.
One of the most compelling scientific motivations to visit Sedna lies in the hypothesis that it may be an exoplanet captured by our solar system during a close stellar flyby with a brown dwarf or solar sibling \cite{Morbidelli_2004, jilkova2015how}. If confirmed, Sedna would represent the first known exoplanetary body accessible to in situ investigation, offering a unique opportunity to directly study extrasolar material. Verifying this hypothesis would require isotopic analyses that can only be conducted on site. In a flyby scenario, preliminary tests of this hypothesis could be performed by collecting and analyzing endogenic volatiles such as methane, water, and ammonia using mass spectrometry or infrared spectroscopy \cite{menten2022endogenically}. Solid silicate particles could also be analyzed using Laser Induced Breakdown Spectroscopy (LIBS), a technique that is advancing rapidly in terms of miniaturization and isotopic selectivity \cite{dellaglio2014libs,zhang2020carbon}. The mere possibility of accessing exoplanetary material at such close range provides strong justification for a dedicated mission.
In summary, a mission to Sedna offers the opportunity to study a pristine, distant object that could shed light on the solar system’s formation, the early history of planetary orbits, and the nature of the farthest reaches of our cosmic neighborhood.

\subsubsection{Missions proposed in literature}
Several studies have analyzed the mission trajectories to Sedna, focusing on optimal launch windows, propulsion methods, and gravity assist maneuvers to minimize travel time and energy requirements. The primary focus of the study \cite{McGranaghan2011} was the design of the interplanetary trajectory for Earth departure dates between 2014 and 2050 to five large trans-Neptunian objects: Quaoar, Sedna, Makemake, Haumea, and Eris. The authors identified Jupiter's gravity assist trajectories that require a total mission $\Delta V$ as low as 7.15 km/s and have arrival V values at the target comparable to those of the New Horizons mission to Pluto (13.8 km/s \cite{NewHorGuo}). The transit time for a mission to reach Sedna was 24.48 years. In 2020 at the 51$^{st}$ Lunar and Planetary Science Conference, authors explored the capability of a solar electric propulsion system to reach Sedna and Eris \cite{BeringElectrical}. They reported that the use of a Variable Specific Impulse Magnetoplasma Rocket (VASIMR) engine propulsion for missions to Aris and Sedna will enable a mission to arrive in half the time required by a chemical propulsion system. It will deliver a spacecraft that is larger and more capable than the one a chemical propulsion system will deliver. 

Considering its perihelion passage is estimated to be in 2073–74, missions to Sedna have been suggested in recent years. In Ref. \cite{Zubko2021} Zubko et al. presented a possible concept including gravity assist maneuvers near Venus, Earth, Jupiter, Saturn and Neptune, with a departure date between 2029-2034. Further details on the optimal trajectory proposed are given in Ref. \cite{Zubko2021_opt}, where Venus, Earth, and Jupiter flybys allow reaching the target in less than 30 years for a $\Delta V$ of less than 5 km s$^{-1}$. The same author investigated the possibility of meeting even more demanding time constraints and limiting the mission to 20 years with a departure in 2041 \cite{Zubko2022}. These studies considered impulsive maneuvers with conventional engines. The options we recommend in this paper would significantly reduce the time required to reach Sedna and are detailed below. The most recent research at the University of Tennessee \cite{Brickley2023} aim was to explore various orbital trajectories to reach the trans-Neptunian object Sedna and find trajectories that could get an orbiter to Sedna as soon as 2070. Trajectories, transit times, arrival speeds, Jupiter flyby distances, and Jupiter radiation doses were compared.

While Sedna’s extreme orbital eccentricity might suggest strict timing constraints, the object remains within a scientifically accessible range for multiple decades; for example, Sedna spends approximately 50 years within 85 AU of the Sun and about 200 years within 100 AU. Given that Sedna moves approximately 1 AU per year at perihelion, this provides a sufficiently long and stable window for mission planning and execution, with even several years of launch delay still allowing for meaningful scientific operations. Mission design must account for orbital inclination and transfer energy, but the perihelion passage provides an extended timeframe for mission feasibility. Jupiter gravity assists or low-thrust propulsion can further optimize transfer windows, ensuring arrival within a timeframe that maximizes science return.

\subsubsection{Main challenges}
Designing a mission to Sedna presents substantial difficulties related to the extreme distance involved, long mission duration, limited propulsion and energy options, and complex navigation required. In this study, we focus on the propulsion system aspects and evaluate options to drastically reduce the time required to reach such a distant target, however when assessing the feasibility of a mission the following challenges must be also taken into account:

\begin{itemize}

    \item Communication Delays: the one-way-light-time (OWLT) for a spacecraft near Sedna would be up to 13 hours at its farthest point from the Sun; for comparison, the OWLT for us to communicate with a satellite orbiting Mars is around 20 minutes. Such a delay would completely exclude the possibility of operating the spacecraft in real time. Communication systems would need to be designed to handle long delays and autonomous systems would be required to execute maneuvers, data collection, and problem-solving with minimal input from Earth.
    \item Navigation and Course Corrections: navigating to a distant and moving target like Sedna is highly complex. Given the vast distances, small errors in trajectory can result in large deviations by the time the spacecraft reaches Sedna. If the mission concept is taking advantage of gravity assists from other planets, the effect of those has to be carefully evaluated and corrected to adjust the spacecraft’s trajectory on the way.
    \item Timing Constraints: when targeting closer objects with orbits similar to Earth's, for a missed launch window opportunity a new one can be identified within months or, in the worst case, a few years. This is clearly not the case for Sedna: its highly elliptical orbit brings it relatively “close” to the Sun (and thus, us) for a small portion of its ~11400-year orbit period.

\end{itemize}

Communicating with a spacecraft at Sedna’s distance poses substantial challenges due to extreme signal attenuation and low bit rates. Drawing from missions like New Horizons and JUICE, the feasibility of data transmission depends on a combination of a high-gain antenna, deep-space ground stations, and sufficient onboard power \cite{stern2015pluto, grasset2013jupiter}. Given Sedna’s distance ($>76$ AU), the mission would likely require an HGA of at least 3m in diameter, with transmission power exceeding that of existing outer solar system missions. While JUICE achieves $>2$ Gb/day from Jupiter, a Sedna mission would face orders-of-magnitude lower data rates ($<<1$ kbps), necessitating highly compressed and prioritized data return. The science return would be constrained by these limitations, requiring careful selection of imaging, spectral, and magnetometer data. Worth mentioning that the huge amount of data could also be collected and then transmitted gradually back to Earth over months and years (strategy similar to that of New Horizons and Europa Clipper missions). Power considerations also play a key role, as deep-space transmission would require significant energy, likely provided by RTGs or a fusion-based system (as for New Horizon, getting to a power generated at launch of roughly 240 W, decreasing to 200 W by Pluto approach in July 2015 \cite{stern2015pluto, stern2015new}). Large ground-based radio telescopes or the DSN’s 70-meter antennas would be essential for data retrieval. A trade-off exists between antenna size, power availability, and payload capacity, all of which impact the overall mission feasibility.

Navigation for a Sedna-bound spacecraft presents unique challenges, including the need for autonomous deep-space guidance. The mission will rely on star trackers for primary attitude determination. Earth-based ranging via the Deep Space Network could enable trajectory correction, while optical navigation using background stellar fields would refine long-range positioning. The mission architecture must balance navigation accuracy with power and mass constraints, ensuring sufficient redundancy without unduly impacting the delivered payload.

Attitude control for a deep-space mission is driven by the need for high-precision pointing for scientific observations and stable alignment for communication. In conventional spacecraft, a combination of reaction wheels and cold-gas thrusters could enable precise control, with momentum management strategies in place to mitigate long-duration drift. 
While for DFD a more conventional AOCS system can be envisioned, attitude control is a critical aspect of solar sail design, as the orientation of the sail directly affects its thrust direction. Conventional spacecraft attitude control methods, such as control moment gyros, reaction wheels, and thrusters, are generally unsuitable for solar sails, especially thrusters, since solar sails are intended to operate without propellant. Ideally, the sail’s center of mass ($c.m.$) aligns with its center of pressure ($c.p.$), but minor manufacturing and deployment errors can create an offset, causing unwanted torque that must be corrected for stable attitude. Conventional attitude control methods cannot effectively maintain the sail’s attitude for several reasons, as first discussed by Wie \cite{Wie2004}. Fortunately, by having an attitude control strategy that allows a controlled $c.m.$-$c.p.$ offset, desired attitude torques can be generated. Fu et al. \cite{Fu2016} give a comprehensive overview of the most common methods for attitude control, for both rigid sailcraft such as control vane, gimbaled/sliding masses, shifted/tilted wings and non-rigid ones. Non-rigid sailcraft rely on internal tension from centrifugal forces and cannot use mechanisms that apply large out-of-plane loads. For non-rigid sails, an attitude control method using reflectivity-controlled membranes has been developed: by adjusting voltage, the membrane's reflectivity changes, shifting the cp and generating a torque for attitude control. Given the relevance of the topic, many papers presented at the latest Space Sailing Symposium (ISSS 2023) focused on attitude control techniques \cite{AnconaKezerashvili2025}.

Besides the ones mentioned above, some additional constraints would need to be considered: the scientific payload needs to be carefully selected, ensuring instruments on board are reliable and capable of operating in cold, dark environments far away from the Sun. The spacecraft would also need to travel through potentially hazardous regions of space, such as areas with high concentrations of cosmic rays or micro meteoroids, thus proper shielding would be required to protect it from impacts and electronics from radiation \cite{NewHorGuo, VoyagerPioneer}. 

The Sedna mission will operate in an extreme deep-space environment, requiring robust thermal management and radiation shielding. Drawing from lessons learned with Voyager and New Horizons, redundancy in critical systems will be necessary to ensure long-term reliability. The spacecraft must incorporate fault detection and autonomous recovery to mitigate the effects of communication delays, while heating elements and radiation-hardened components will safeguard operational stability beyond Neptune \cite{stern2015new}.

These are some of the major adversities and risks such a mission would face. In the following sections, we will restrict our analysis to propulsion technologies and mission concepts, however, any of the proposed solutions, when chosen, would require some iterations in the design phases to overcome these challenges.

\section{The need for innovative propulsion concepts}

Due to the limitations of traditional methods, innovative propulsion systems are crucial to reach distant targets like Sedna. Chemical propulsion, while providing high thrust for launches, suffers from low efficiency and high fuel mass requirements for long-duration missions: the Voyager 1 spacecraft, launched in 1977, traveled with a cruise speed of 17 km/s (3.57 AU per year). Electric propulsion, including ion and Hall effect thrusters, offers much higher efficiency and finds many applications nowadays \cite{Bhatia2022} but produces insufficient thrust for rapid deep-space travel. NASA's ion-thrusters-equipped Dawn spacecraft, which explored the asteroid belt, highlighted the advantages of electric propulsion for deep-space maneuvers \cite{RAYMAN2022}, however both Vesta and Ceres were relatively close (1.14 AU and 2.87 AU) when reached by Dawn in 2011 and 2015 respectively.
It is evident that novel space propulsion mechanisms are essential for enabling faster and more energy-efficient exploration of the outer solar system and beyond. In the following Subsections, we discuss two alternatives to overcome these challenges: a thermonuclear fusion engine, the Direct Fusion Drive,  and solar sails, enhanced by thermal desorption of coatings. 

\subsection{Direct Fusion Drive}

The Direct Fusion Drive is a D-$^{3}$He-fueled, aneutronic, thermonuclear fusion propulsion system that is under development at Princeton University Plasma Physics Laboratory \cite{Cohen2019}. In previous collaborations, the authors of this study have identified potential applications for this breakthrough technology. With a power range of 1 to 10 MW, depending on mission-specific requirements, the DFD demonstrates versatility for a variety of use cases. While a broad overview of its capabilities to provide the required mobility even for human planetary exploration is given in Ref. \cite{Genta2020}, analysis of realistic trajectories for robotic missions to Titan \cite{Gajeri2021}, trans-Neptunian objects \cite{Aime2021} and other destinations in the solar system \cite{KezerashviliDFD2021} have been performed.

\begin{figure*}[h]
     \centering
     \includegraphics[width=0.6\linewidth]{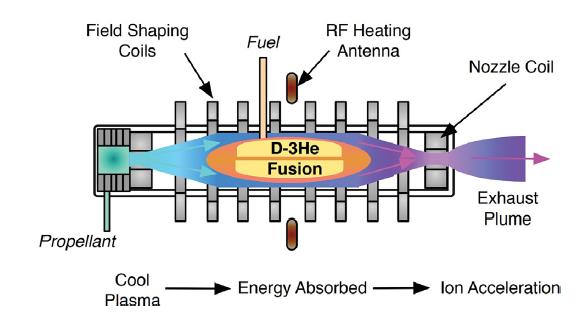}
     \caption{Schematic diagram of the Direct Fusion Drive engine subsystems with its simple linear configuration and directed exhaust stream. A propellant is added to the gas box. Fusion occurs in the closed-field-line region. Cool plasma flows around the fusion region, absorbs energy from the fusion products, and is then accelerated by a magnetic nozzle.
Credits \cite{Cohen2019}.}
     \label{fig:DFD}
\end{figure*}

\subsubsection{Engine characterization} 
The Direct Fusion Drive is a groundbreaking propulsion system that produces both propulsion and electrical power from a compact fusion reactor. Funded by NASA and based on the Princeton Field Reversed Configuration (PFRC) fusion experiment \cite{Cohen2000,Cohen2007}, DFD is designed for various space missions, including manned and robotic missions to Mars, and cargo missions to the outer solar system. The PFRC-2 uses a unique radio frequency plasma heating method, known as "odd-parity heating," to increase plasma temperatures for fusion \cite{Cohen2000, Cohen2000a}. The technique relies on a magnetic field structure that maintains plasma confinement, which is crucial for achieving optimal fusion conditions. This configuration ensures stable plasma behavior, making it more resilient to instability than other fusion devices.

A diagram of the DFD is provided in Fig. \ref{fig:DFD} and characteristics for low and high power configurations of the engine are listed in Table \ref{tab:DFD_config} below.

\begin{table}[h]
\caption{DFD characteristics for low and high power configurations \cite{Cohen2019}.}
\label{tab:DFD_config}
\centering
\begin{tabular}{lll}

                         & Low power   & High power \\ \hline
Fusion Power (MW)        & 1           & 10          \\ 
Specific Impulse (s)     & 8500-8000   & 12000-9900  \\
Thrust (N)               & 4-5      & 35-55        \\ 
\end{tabular}
\end{table}

According to Ref. \cite{Cohen2019}, the current estimated DFD specific powers are between 0.3 and 1.5 kW/kg and a shielding thickness of 22 cm would result in superconducting coil's lifetime of 13 years.

\subsubsection{Technical challenges}
The selection of fuel for DFD is one of the key design aspects. The fusion reaction of nuclei of deuterium (D), and tritium (T) is the most promising for the energy produced. However, D–T fuel is not ideal for the DFD, due to the significant emission of neutrons. When neutrons are generated in fusion reactions, part of the energy becomes unusable, causing energy losses and neutron emissions that can contaminate the spacecraft. To protect the spacecraft (and the crew, if present) shielding is necessary since neutrons, being uncharged, cannot be controlled by electric or magnetic fields. Reducing neutron flux is crucial to minimize material damage and radiation but would also increase the mass needed for radiation shielding. For these reasons, aneutronic fuel, such as a deuterium and $^{3}$He mixture, is preferred. This choice addresses the issue of neutron production, however the DFD - like every cutting-edge technology - holds avenues for improvement.
One of the main limitations to be accounted for is that, currently, only about 30 kg of $^{3}$He is available on Earth, so it would need to be sourced through mining, either from terrestrial or off-world resources, or by synthesizing it. It has been estimated that the full budget of fission reactors production worldwide could provide no more than 110 kg of $^{3}$He per year under the unlikely hypothesis that it could all be separated from the rest, while the totality of natural gas wells could provide less than 5 kg/yr \cite{wittenberg1992review}. Tables \ref{tab:fusion_comparison1} and \ref{tab:fusion_comparison2} summarize the main characteristics for fusion reactions.

\begin{table}[h]
    \caption{Comparison of Nuclear Fusion Reactions: ignition temperatures, energy yield per reaction and neutron production.}
    \vspace{0.2cm}
    \label{tab:fusion_comparison1}
    \centering
    \begin{tabular}{cccc}
        \textbf{Reaction} & \textbf{Ignit. T (MK)} & \textbf{Energy (MeV)} & \textbf{Neutron Prod.} \\
        D-D   & $\sim$300 & 3.3–4.6  & Moderate \\
        D-T   & $\sim$150 & 17.6  & High \\
        D-$^3$He & $\sim$500 & 18.3  & None (direct) \\
    \end{tabular}
\end{table}

\begin{table}[h]
    \caption{Comparison of Nuclear Fusion Reactions: fuel availability and main challenges.}
    \vspace{0.2cm}
    \label{tab:fusion_comparison2}
    \centering
    \begin{tabular}{cccccc}
        \textbf{Reaction} & \textbf{Fuel Availability} & \textbf{Main Challenges} \\
        D-D   & High (seawater) & Lower efficiency, neutron radiation \\
        D-T   & Low & Neutron damage, tritium scarcity \\
        D-$^3$He & Extremely rare & High ignition temp, fuel scarcity \\
    \end{tabular}
\end{table}

The D-$^{3}$He reaction is appealing since it has a high energy release and produces only charged particles (making it aneutronic). This allows for easier energy containment within the reactor and avoids the neutron production associated with the D-T reaction. However, the D-$^{3}$He reaction faces the challenge of a higher Coulomb barrier, requiring a reactor temperature approximately six times greater than that of a D-T reactor to achieve a comparable reaction rate.

A detailed review of the initial DFD concept and suggestions for modifications were provided by Jain et al. \cite{Jain2023}. For example, replacing the conventional turbine and radiator system with a magnetohydrodynamic (MHD) generator could reduce complexity, mass, and maintenance requirements on long deep-space missions. Further considerations and ongoing research on this topic are available online, we won't expand on the subject that is beyond the scope of this work.

\subsection{Solar Sails}
Scientists and writers have been dreaming of solar sailing for a long time; interest in solar sails grew in the 1970s with NASA's theoretical studies, though no practical missions materialized at the time. 
The first attempt to showcase solar sailing was the Cosmos-1 mission, which failed due to a launch vehicle malfunction in 2005 \cite{wright1992space}. Japan's IKAROS mission, launched by JAXA in 2010 \cite{Tsuda2011,Kezer2011,Tsuda2013}, became the first spacecraft to employ solar sail propulsion in interplanetary space successfully. The sail generated measurable thrust, proving that sunlight could effectively power spacecraft for deep-space missions \cite{mcinnes1999solar}. In parallel, NASA’s NanoSail-D2 mission \cite{NanoD2011}, launched the same year, validated the viability of solar sailing in Earth orbit on a smaller scale. These early milestones paved the way for more advanced applications. In 2015 LightSail-1 successfully deployed its sail \cite{Biddy2012} and in 2019 LightSail-2 became the first small spacecraft to change its orbit using only sunlight for propulsion \cite{planetary2019lightsail2}. This success of the Planetary Society's sails highlighted their potential for propelling small, fuel-less spacecraft, critical for long-duration missions in deep-space. Ongoing research continues to focus on improving sail materials and deployment mechanisms for future interplanetary missions. The most recent achievement in the sector concerns the NASA Advanced Composite Solar Sail System (ACS3) mission: ACS3 is a technology demonstration of solar sail technology for future small spacecraft, consisting of a 12U CubeSat small satellite (23 cm x 23 cm x 34 cm; 16 kg) that unfolds a quadratic 80 m$^2$ solar sail. After its launch in April, it was confirmed as successfully operational on August the 29th, 2024 \cite{ACS3news}. Although the basic idea behind solar sailing appears simple, challenging engineering problems have to be solved \cite{LEIPOLD2003}. The Solar Sailing community is actively making an effort to address these by fostering collaboration and sharing updates on any progress regularly, also in the format of an International Symposium, such as the 6th International Symposium on Space Sailing (ISSS) that was held in 2023 at the CUNY New York City College of Technology \cite{ancona2025recent}.

\subsubsection{Thermal desorption providing additional thrust}
\begin{figure}[h]
     \centering
     \includegraphics[width=0.6\linewidth]{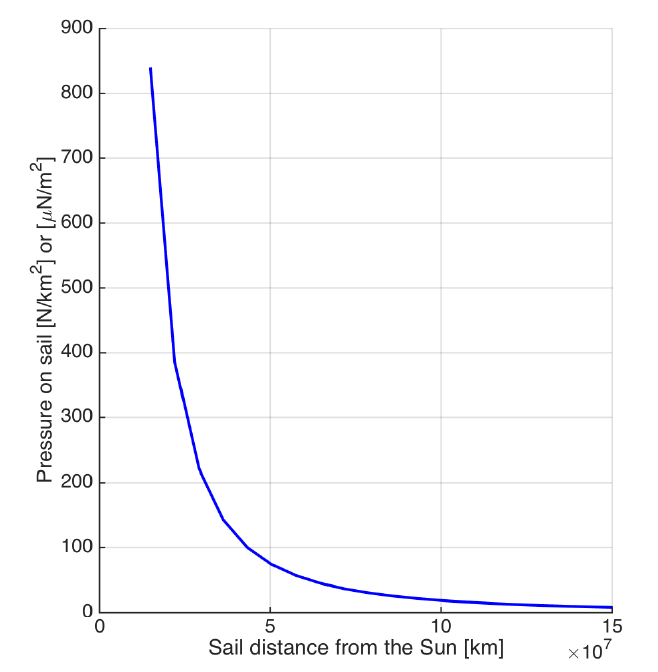}
     \caption{Pressure on the sail as a function of the sail distance from the Sun, assuming reflectivity of 0.85.}
     \label{fig:Press_sail}
\end{figure}

Exploring the outer solar system using solar sail propulsion requires achieving high cruise speeds, which can be facilitated by accelerating the sailcraft in the region near the Sun. Solar sails are one of the few propulsion methods with great potential, as they rely on the Sun's electromagnetic radiation for thrust and do not require onboard fuel. The efficiency of a solar sail increases as it approaches the Sun because of the stronger solar radiation pressure, as shown in Fig. \ref{fig:Press_sail}.

To evaluate the pressure exerted on the sail, it is necessary to consider
the momentum transported not by a single photon, but by a flux of
photons. Based on the inverse square law the energy flux $\phi$ (also
said solar irradiance) depends on the distance $r$ of the body from
the Sun as: 

\begin{equation}
\phi\left[\frac{W}{m^{2}}\right]=\phi_{E}\left(\frac{R_{E}}{r}\right)^{2}=\frac{3.04\cdot10^{25}}{r^{2}},\label{eq:flux}
\end{equation}

where $R_{E}=1\:AU$ (Sun-Earth distance) and $\phi_{E}=L_{S}/\left(4\pi R_{E}^{2}\right)=1346\:W/m^{2}$
is the Solar irradiance at Earth distance, defined through the solar
luminosity $L_{S}$. The pressure on the sail can then be written as function of the reflectivity parameter $\varrho$:

\begin{equation}
P_{sail}=\frac{1+\varrho}{c}\phi.\label{eq:psail}
\end{equation}

Deploying a solar sail close to the Sun maximizes the force exerted by the radiation, enabling higher acceleration and faster cruise speeds. To further enhance this acceleration, we propose utilizing space environmental effects; by coating the solar sail with materials that undergo desorption at elevated temperatures, we can introduce an additional acceleration mechanism \cite{Kezer2015,Ancona2019}.
The novel thermal desorption propulsion method of solar sail was suggested in Refs. \cite{Benford2005,Benford2006} consists of using a microwave beam to heat a solar sail coating by substance until its surface coat sublimes or desorps. This process can add a significant thrust in addition to the photon thrust. The microwave methods of heating were considered using a ground-based stationary microwave beamer or the orbiting microwave beamer behind a solar sail in the same initial circular orbit. In our proposal, the heating comes directly from the Sun.
As the sail heats up from solar radiation at a specific heliocentric distance, the coated material desorbs, releasing additional energy that boosts the sail's speed \cite{Ancona2019_KB,KezStar2021}. This effect, combined with the traditional solar radiation pressure, provides an extra source of propulsion. The perihelion of the solar sail's orbit is selected based on the temperature required for the desorption of the coating material. At this critical perihelion, the sail, that was stowed, is deployed and reaches its maximum temperature, triggering desorption and subsequent acceleration. The sail achieves escape velocity from this process but continues to accelerate further due to the radiation pressure. By leveraging both the increased solar radiation pressure near the Sun and the additional thrust from material desorption, this approach offers a promising method for reaching high velocities, making it highly suitable for missions to the outer solar system and beyond.

From a chemical perspective, we refer to the thermal desorption process as follows: an "adatom" (adsorbed atom) present on a surface at low temperatures may remain almost indefinitely in that state. As the temperature of the substrate is increased, however, there will come a point at which the thermal energy of the adsorbed species is such that it may desorb from the surface and return into the gas phase.
The rate of desorption $R_{d}$ of an adsorbed species from a surface
can be expressed in the general form as:

\begin{equation}
R_{d}=k_{d}\left(N_{A}\right)^{q},\label{eq:ratede}
\end{equation}

\noindent where $N_{A}$ is the surface concentration of adsorbed species, $q$ is the kinetic order of desorption and $k_{d}$ is the rate constant for desorption. The latter one is commonly described as:

\begin{equation}
k_{d}=\nu_{0}\:\exp\left(-\frac{E_{A}}{k_{B}\mathcal{T}}\right),\label{eq:costde}
\end{equation}

\noindent where $\nu_{0}$ is the pre-exponential frequency factor and its typical value is $10^{13}s^{-1}$, $E_{A}$ is the activation or liberation energy (usually $\le1\,eV$), $\mathcal{T}$ is the temperature and $k_{B}=1.38\cdot10^{-23}\,JK^{-1}$ is the Boltzmann constant. 

The order of desorption $q$ mostly depends on the type of considered
reaction: usually it is a first-order process if it involves atomic
or simple molecular desorption. Sometimes $\nu_{0}$ is also called
the "attempt frequency" at overcoming the barrier to desorption.
In the particular case of simple molecular adsorption, $\nu_{0}$
corresponds to the frequency of vibration of the bond between the
atom or molecule and substrate. This is because every time the bond
is stretched during the course of a vibrational cycle could be considered
as an attempt to break the bond and, hence, an attempt at desorption.
In general, it was found that the pre-exponential frequency factor
can span a wide range of values from $10^{13}s^{-1}$ to $10^{21}s^{-1}$
depending on the vibrational degrees of freedom of the adsorbate.
From Eqs. (\ref{eq:ratede}) and (\ref{eq:costde}) the general expression
for the rate of desorption can be obtained, considering that it corresponds
to the time reduction of adatoms present on the surface:

\begin{equation}
R_{d}=-\frac{dN_{A}}{dt}=\nu_{0}\:\left(N_{A}\right)^{q}\:\exp\left(-\frac{E_{A}}{k_{B}\mathcal{T}}\right).\label{eq:costde-1}
\end{equation}

The rate of mass loss under heating $dN_{A}/dt$ is the desorbed flux
in $atoms/m^{2}s$. Note that the exponential factor suggests that
the desorption will have a sudden onset after the surface gets warm
\cite{Benford2005}. Equation (\ref{eq:costde-1}) can be formally
solved when temperature varies with time. As the activation energy
increases, the time to desorb gets longer, because the rate of desorption
decreases.

Considering a sail coated with a material that undergoes thermal
desorption, possibly Carbon, as suggested in Ref. \cite{Benford2005}.
For the common graphite, Carbon $^{12}C$, the molecular mass is $m_{p}=19.2\cdot10^{-27}\:kg$
and its density is $d=2267\:kg/m^{3}$, whereas Fullerene $C_{60}$
has $m_{p}=1203.5\cdot10^{-27}\:kg$ and its density is
$d=1650\:kg/m^{3}$. Once the material is known,
one can find the number of atoms desorbed and then the number of atoms desorbed per unit area during the time, $dN_{A}/dt$, is obtained from (\ref{eq:costde-1}). If the sail area is defined, $N_{A}$ is fixed, so the required desorption time can be found.
However this approach would require very accurate information about
the material's characteristics.

\subsubsection{Material degradation due to close proximity to the Sun}
A key challenge in a solar sailing mission concept like the one proposed is the potential degradation of sail materials due to high temperatures during the perihelion passage. As the spacecraft approaches the Sun, intense solar radiation can lead to structural and optical degradation through desorption effects, mechanical stress, and material outgassing. The authors of this paper have previously addressed these concerns in Ref. \cite{Ancona2017_temp, Kezer_temp, Kezer_book} studies on the behaviour of solar sail material in extreme thermal environments which examine the effects of temperature on material stability and reflectivity. Specifically, these analyses include:

1) Temperature dependence of optical properties: Optical degradation of sail materials at high temperatures can alter mission performance. We consider temperature-dependent emissivity, which leads to a more gradual increase in sail temperature compared to the case of constant emissivity.

2) Thermal modeling for perihelion passage: Our calculations indicate that, when considering realistic emissivity variations, the sail can withstand higher temperatures than previously assumed \cite{Ancona2017_temp}. This allows the spacecraft to approach closer to the Sun without significant degradation of optical properties.

Solar wind effects were also considered, and the induced sail erosion for different materials. Specifically, the interaction of several materials made of light atoms have been investigated on the basis of physical considerations \cite{kezerashvili2008solar,matloffmerit}. These results suggest good resilience. The layer of Carbon is expected to desorb during short time, coherently with its function as booster associated with desorption.

By incorporating these factors, we provide a more accurate assessment of sail survivability in the extreme thermal environment near the Sun, reinforcing the feasibility of the proposed mission architecture.

\section{Proposed Mission Concepts}
Following the description of the propulsion mechanisms considered, this section provides details on the suggested mission profiles for both the DFD and the solar sail. The results presented below have been obtained in
various study phases and with different applications. Besides from classic Mission Analysis tools, such as Ansys STK and NASA GMAT, ad-hoc scripts were realized for specific purposes (Matlab, python, etc.).

\subsection{DFD Scenario}
For a trajectory design to trans-Neptunian object Sedna, we considered a 1.6 MW DFD, assuming constant thrust and specific impulse. Within a Thrust-Coast-Thrust (TCT) profile 3 phases are identified: the Earth escape spiral trajectory; the interplanetary travel coasting phase; and final maneuvers to rendezvous with the target. 
As demonstrated in \cite{KezerashviliDFD2021} a spacecraft with DFD would reach the TNO Eris (at a distance of 78 AU at rendezvous) in 10 years. Considering this matches with Sedna's perihelion, DFD would decrease the time required by 50\% (compared to the 20 years of the traditional propulsion systems in Ref. \cite{Zubko2022}).

\begin{figure}[h]
     \centering
     \includegraphics[width=0.8\linewidth]{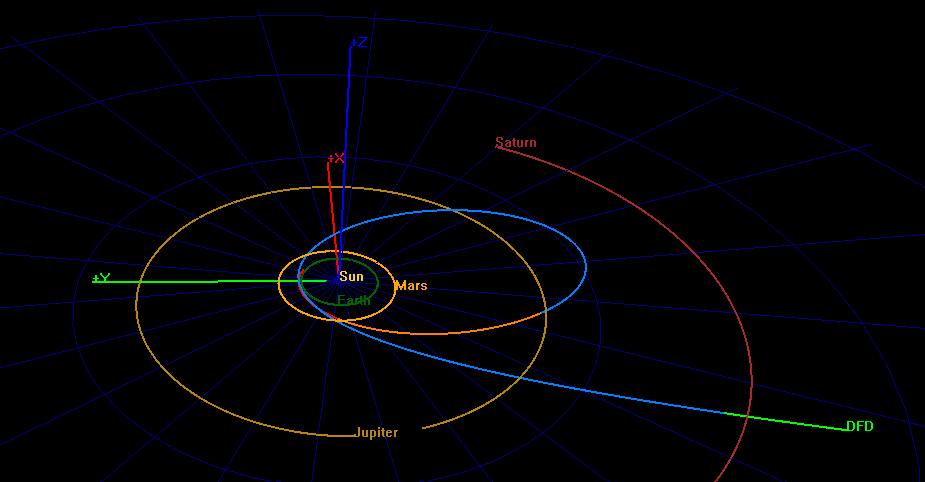}
     \caption{DFD on an escape trajectory (obtained with GMAT tool).}
     \label{fig:DFD_Escape}
\end{figure}

A valid solution to reduce the time even more would be to evaluate a constant thrust (CT) profile; thus the disadvantage is that this would require additional propellant, and result in a 1/3 reduction of payload mass. In our analysis, the goal is to deliver at least 1500 kg of payload to the destination in under 10 years. With a specific power of 0.8 kW/kg, the DFD engine weighs 2000 kg, resulting in a total dry mass of 3500 kg. Based on initial estimates, approximately 4000 kg of propellant would be required, bringing the total launch mass to 7500 kg. It is important to note that these values would need to be refined during a preliminary study, once the mission’s scientific objectives and requirements have been defined. This would also allow for a more accurate assessment of the spacecraft's configuration, structural design, and corresponding mass.  The total $\Delta$V for the mission reaches 80 km/s, with half of that needed to slow down during the rendezvous phase, where the coasting velocity is 38 km/s. Each maneuver would take between 250 and 300 days, requiring about 1.5 years of thrust over the 10-year journey. However, the engine would remain active to supply power to the system. The amount of $^{3}$He required is estimated at 0.300 kg. Note that in this preliminary study, no flyby opportunity was considered; including one or more gravity assists could significantly decrease the necessary $\Delta$V, as thoroughly analyzed in Ref. \cite{Zubko2021_opt}.
We considered launch opportunities starting from 2047, as that would allow for significant technological progress and still leave time for the preparation of mission execution. 

\begin{table}[h]
\caption{Summary of DFD Mission Parameters for a mission to Sedna considering TCT profile. The total thrust duration is accounting for the 10-year mission.}
\vspace{0.2cm}
\centering
\begin{tabular}{|p{6cm}|p{3cm}|}
\hline
Fusion Power & 1.6 MW \\
\hline
Specific Impulse & 9600 s \\
\hline
Thrust & 8 N \\
\hline
Specific Power & 0.8 kW/kg \\
\hline
DFD Engine Mass & 2000 kg \\
\hline
S/C Mass (Dry) & 3500 kg \\
\hline
Propellant (D) Mass & 4000 kg \\
\hline
Launch Mass (Wet) & 7500 kg \\
\hline
Total $\Delta V$ & 80 km/s \\
\hline
Coasting Velocity & 38 km/s \\
\hline
Thrust Duration per Maneuver & 250–300 days \\
\hline
Total Thrust Duration & $\sim$1.5 years \\
\hline
$^{3}$He Required & 0.300 kg \\
\hline
\end{tabular}
\label{tab:dfd_summary}
\end{table}

A final consideration regarding the proposed strategy is that the TCT profile mission assumes the DFD can be switched on and off for thrust generation. This is a crucial assumption, as it requires the engine to remain inactive for approximately a year during the coasting phase. While theoretically feasible, this capability has not yet been confirmed.

\subsection{Enhanced Solar Sail Scenario}

For what concerns the solar sail, it consists of a reflective membrane coated by materials that undergo thermal desorption attached to an inflatable torus-shaped rim \cite{KezStar2021}.
The sail's deployment from its stowed configuration is initiated by introducing inflation pressure into the toroidal rim with an attached circular flat membrane coated by heat-sensitive materials that undergo thermal desorption at the perihelion of the heliocentric escape orbit and provide an acceleration - in addition to the conventional one due to solar electromagnetic radiation. We suggest the following sequence for the mission: elliptical transfer, gravitational slingshot, followed by thermal desorption. Initially, the spacecraft performs a transfer from Earth’s orbit to Jupiter’s orbit using conventional propulsion. A gravitational assist from Jupiter then redirects the trajectory towards a close approach to the Sun. At the perihelion, the solar sail is deployed, and its coating undergoes desorption. This results in two sources of acceleration: one from the material desorption of the coating and the other from the solar radiation pressure, propelling the spacecraft toward deep-space. The perihelion distance to aim for depends on the coating material.

A flyby, also known as a gravity assist or slingshot maneuver, is a technique used to modify a spacecraft’s trajectory and speed by passing close to a planet. Instead of relying solely on onboard propulsion, the spacecraft harnesses the planet's gravitational field to gain or lose energy relative to the Sun. This method allows interplanetary missions to reach distant destinations with reduced fuel consumption, making exploration beyond the inner Solar System more feasible.
The concept was first developed in the 1960s at NASA’s Jet Propulsion Laboratory (JPL) and was successfully demonstrated by Pioneer 10 in 1973. By passing behind Jupiter, the spacecraft gained a significant velocity boost, enabling it to escape the Solar System. Since then, gravity assists have been fundamental to nearly all interplanetary missions. In our proposed concept, the sail (still folded) would reach Jupiter, where the flyby occurs.

\begin{figure}[h]
     \centering
     \includegraphics[width=0.9\linewidth]{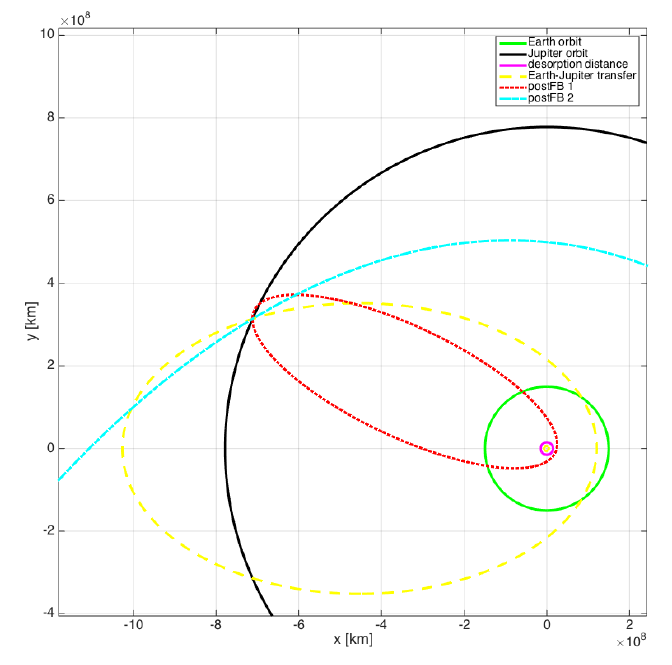}
     \caption{Jupiter Flyby: increase and decrease in orbital energy when passing behind (cyan orbit, dash-dot line) or in front (red orbit, dotted line) of the planet, respectively.}
     \label{fig:Flyby_sail}
\end{figure}

\begin{figure}[h]
     \centering
     \includegraphics[width=0.9\linewidth]{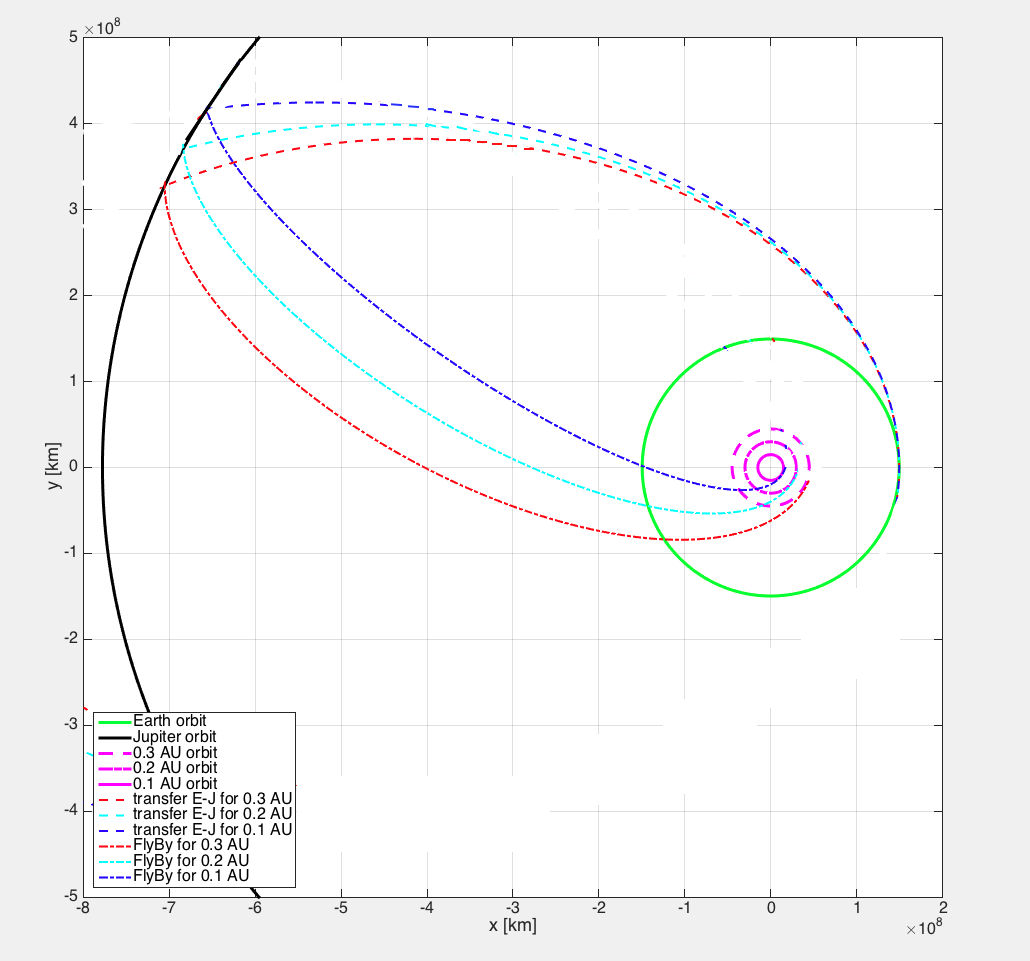}
     \caption{Elliptical transfer options from Earth to Jupiter and Flyby to the inner orbit, for heliocentric target distances of 0.3, 0.2 and 0.1 AU respectively.}
     \label{fig:scen2}
\end{figure}

\begin{table}[h]
\caption{Earth - Jupiter heliocentric transfer, Jupiter flyby and post-flyby orbit details for the different heliocentric distances considered. The time of flight refers to the whole sequence: from leaving Earth to the moment where the target perihelion is reached, before the sail is deployed (that is taken into account in the following table). $r_{E}$ and $r_{J}$ are Earth and Jupiter heliocentric distances respectively.}
\vspace{0.2cm}

\begin{centering}
\begin{tabular}{llrrrrrrr}
\hline 
radius of perihelion, $r_{P}$ & $\left[AU\right]$ &  &  & 0.3 &  & 0.2 &  & 0.1\tabularnewline
temperature, $\mathcal{T}$ & $\left[K\right]$ &  &  & 737 &  & 865 &  & 1140\tabularnewline
specific energy, $E_{gt}$ & $\left[J/kg\right]$ &  &  & -117 &  & -109 &  & -98\tabularnewline
speed at $r_{E}$, $v_{1}$ & $\left[km/s\right]$ &  &  & 39.20 &  & 39.41 &  & 39.70\tabularnewline
speed at $r_{J}$, $v_{2}$ & $\left[km/s\right]$ &  &  & 10.29 &  & 11.08 &  & 12.06\tabularnewline
hyp. ex. vel. at Earth, $v_{\infty_{e}}$ & $\left[km/s\right]$ &  &  & 9.44 &  & 9.65 &  & 9.94\tabularnewline
$\Delta V$ required, $\Delta V_{0}$ & $\left[km/s\right]$ &  &  & 6.72 &  & 6.86 &  & 7.05\tabularnewline
hyp. ex. vel. at Jupiter, $v_{\infty_{t}}$ & $\left[km/s\right]$ &  &  & 8.91 &  & 9.76 &  & 10.79\tabularnewline
rotation angle, $\alpha$ & $\left[deg\right]$ &  &  & 142 &  & 138 &  & 134\tabularnewline
``free'' $\Delta V$ gain, $\Delta V_{fb}$ & $\left[km/s\right]$ &  &  & 16.84 &  & 18.24 &  & 19.88\tabularnewline
ang. between $v_{t}$ and $v_{\infty_{t}}^{-}$ & $\left[deg\right]$ &  &  & 52 &  & 56 &  & 60\tabularnewline
speed after flyby, $v_{4}$ & $\left[km/s\right]$ &  &  & 4.86 &  & 4.32 &  & 3.69\tabularnewline
specific energy, $E_{g\:post}$ & $\left[J/kg\right]$ &  &  & -158 &  & -161 &  & -163\tabularnewline
speed at $r_{P}$, $v_{5}$ & $\left[km/s\right]$ &  &  & 73.24 &  & 91.28 &  & 129.11\tabularnewline
time of flight, $t_{f}$ & $\left[years\right]$ &  &  & 3.71 &  & 3.58 &  & 3.01\tabularnewline

\hline 
\end{tabular}
\par\end{centering}
\label{scen2tab}
\end{table}

\begin{table}[h]
\caption{Sail cruise speed and distance covered considering thermal desorption.}
\vspace{0.2cm}
\begin{centering}
\begin{tabular}{llrrrrrr}
\hline 
radius of perihelion, $r_{P}$ & $\left[AU\right]$  & 0.3 &  & 0.2 &  & 0.1\tabularnewline
speed at $r_{P}$ before desorption & $\left[km/s\right]$  & 73.24 &  & 91.28 &  & 129.11\tabularnewline
speed at $r_{P}$ after desorption & $\left[km/s\right]$   & 181.82 &  & 226.90 &  & 321.44\tabularnewline
cruise speed, $v_{cruise}$ & $\left[km/s\right]$ & 185.04 &  & 230.76 &  & 326.90\tabularnewline
distance/year, $AU_{y}$ & $\left[AU/year\right]$ & 39.0 &  & 48.7 &  & 68.9\tabularnewline
\hline 
\end{tabular}
\par\end{centering}
\label{tab2res}
\end{table}

Note that, as shown in Fig. \ref{fig:Flyby_sail}, a flyby can either reduce or increase the spacecraft velocity, depending on its approach: if the spacecraft is passing ``ahead'' of the planet, it will result in a
decrease of the final heliocentric velocity that
corresponds to reducing the orbit energy. Differently,
if the spacecraft is passing ``behind'' the planet, this will result in a
increase of the final heliocentric velocity that
corresponds to higher orbit energy. For example, as shown in Fig. \ref{fig:Flyby_sail}, the Jupiter flyby could change the yellow (dashed line) transfer orbit from Earth to Jupiter into the red (dotted line) orbit if passing ``ahead'' of the planet, or into the cyan blue (dash-dot line) orbit if the spacecraft passes behind the planet. In this particular case, since our goal is to get closer to the Sun, we are considering a reduction of orbital energy (red line).

Fig. \ref{fig:scen2} depicts inbound trajectories for different target heliocentric distances. Specifically, it illustrates cases where the spacecraft aims to reach 0.3 AU, 0.2 AU, and 0.1 AU after a Jupiter flyby. The most significant variables for the sequence of events described in the scenario are presented in Table \ref{scen2tab}. This comparison highlights how the initial conditions of the flyby influence the achievable post-flyby trajectory and the feasibility of reaching extreme inner Solar System destinations.
Once reached the target perihelion distance, the sail is deployed. Before introducing the desorption contribution, we evaluate what performances we could obtain with a conventional solar sail.
In the best case, thus when reaching the closest heliocentric distance of 0.1 AU, the speed achieved at $r_{P}$ is 129.11 km/s from Tab. \ref{scen2tab}. The resulting cruise velocity, for a $\beta$ = 0.75, would be 106.9 km / s, which would allow travelling 22.6 AU per year.

Results of calculations for the velocity gain due to thermal desorption
acceleration for the three different perihelion radii presented in Table \ref{tab2res} refer to a high-performance sail, with $\beta$ = 1.2 and sail loading factor $\sigma$ = 1 g/m$^{2}$.
When we assume more realistic values, the following parameters are considered for the sail: lightness number $\beta$ = 0.75, areal mass $\sigma$ = 5 g/m$^{2}$, coating mass M$_{0}$ = 1 kg, mass of payload M$_{P}$ = 1.5 kg, desorption rate m$_{0}$ = 1 g/s. For a perihelion of 0.3 AU, the resulting cruise speed would reach about 100 km/s. Traveling at about 20 AU/year, from the moment of deployment at the perihelion it would take approximately 4 years to arrive at Sedna; however, the time for the transfer to Jupiter first and then the desorption target distance has to be added, giving a total mission time of 7 years.

\section{Conclusions}
This study explores two promising propulsion technologies, the Direct Fusion Drive and solar sails enhanced by thermal desorption, for deep-space missions targeting the trans-Neptunian object Sedna. Both approaches aim to overcome the significant challenges posed by Sedna’s vast distance and the long duration of such missions. We specifically examine two mission architectures for a Sedna flyby: one utilizing an augmented solar sail and another leveraging DFD propulsion. These approaches offer fundamentally different trade-offs in terms of payload capacity, mission duration, and scientific return.

\begin{table}[h]
\centering
\caption{Comparison of DFD and Solar Sail Scenarios for a Mission to Sedna}
\vspace{0.2cm}
\begin{tabular}{|p{2cm}|p{4cm}|p{4cm}|}
\hline
\textbf{} & \textbf{DFD} & \textbf{Enhanced Solar Sail} \\
\hline
\textbf{Mission} & Robotic exploration with comprehensive scientific instruments & Preliminary scouting mission with minimal instrumentation \\
\hline
\textbf{Time} & $\sim$10 years & $\sim$7 years \\
\hline
\textbf{Payload} & Up to 1500 kg & $\sim$1.5 kg \\
\hline
\textbf{Propulsion} & high-efficiency nuclear fusion engine & Photon-driven propulsion with initial boost from thermal desorption \\
\hline
\textbf{Challenges} & Requires development and validation of advanced fusion technology; reactor shielding & Very low payload mass limits onboard systems; trajectory control and thermal material limits near the Sun \\
\hline
\textbf{Science} & High: enables full suite of instruments for imaging, spectroscopy, in-situ sampling & Limited: suitable for basic measurements and initial characterization \\
\hline
\textbf{TRL} & Low to medium; under research and prototyping & Medium; demonstrated components, but desorption methods still experimental \\
\hline
\textbf{Suitability} & Best for full-scale exploration and sample-return scenarios & Ideal for precursor missions and tech demos in deep-space \\
\hline
\end{tabular}
\label{tab:dfd_vs_sail}
\end{table}

The DFD, with its high efficiency and power, is ideally suited for robotic exploration missions, enabling substantial payloads and comprehensive scientific analysis. Among the most cutting-edge propulsion technologies, the DFD would be a true game-changer for any robotic mission to distant targets, offering a payload capacity in excess of 100 kg, and potentially up to 1500 kg. This enables a broader suite of scientific instruments, including full-scale mass spectrometers for in situ composition analysis, radar or other systems to investigate surface topology, and high-resolution imaging equipment for detailed surface mapping.

On the other hand, the solar sail mission, while constrained to a small payload (~1.5 kg), benefits from ongoing advancements in miniaturized technology that open new opportunities for deep-space exploration. Inspired by initiatives such as Breakthrough Starshot, the integration of compact, low-mass scientific instruments can enable meaningful data collection within strict mass constraints. Potential scientific payloads could include miniaturized spectrometers for surface composition analysis, compact magnetometers to study Sedna’s magnetic environment, and high-sensitivity cameras for imaging in the visible and infrared spectra. This mission profile, although limited in terms of in situ capabilities, could still yield valuable remote sensing and spectroscopic data.

Different mission profiles will naturally yield different types and depths of science return. The DFD scenario supports more comprehensive investigations, while the sail-based mission could serve as a valuable precursor or reconnaissance mission.

This study highlights the feasibility and trade-offs of two distinct mission concepts to explore Sedna. While the augmented solar sail offers a novel approach with potentially shorter transit times, it comes with stringent constraints on payload mass and communication capabilities. In contrast, the DFD architecture enables a significantly larger payload and extended mission lifetime, although at the cost of longer travel durations and higher mission complexity. Both concepts require careful planning around launch and arrival windows, trajectory design, and system reliability.

We acknowledge the challenges associated with deep-space communication, power generation, and long-duration spacecraft reliability. Nevertheless, advancements in propulsion and miniaturization technologies suggest that both mission scenarios could provide meaningful contributions to our understanding of the outer solar system.

In summary, Sedna’s discovery has broadened our understanding of the solar system’s outermost regions, raising fundamental questions about planetary formation and the existence of distant gravitational influences. While no dedicated missions have yet targeted Sedna, its scientific importance remains compelling. Given that proposed launch windows fall within the planning horizon of current space programs, further development of advanced propulsion systems—including both solar sail augmentation and fusion-based solutions—will be critical to unlocking the next era of deep-space exploration.

\section*{References}

\printbibliography[heading=none]

\end{document}